\documentclass[aps,prb,twocolumn,superscriptaddress]{revtex4-1}

\usepackage{natbib}
\bibpunct{(}{)}{;}{a}{}{,}
\usepackage{makecell}
\usepackage[pagebackref=true]{hyperref}

\usepackage{graphicx}
\usepackage{bm}
\usepackage{amsmath}
\usepackage{amssymb}
\usepackage{hyperref}
\usepackage{mhchem}
\usepackage{color}

\begin{document}

\title{Density Functional Theory Calculations on the Interstellar Formation of Biomolecules}

\author{Qingli Liao}
\affiliation{Laboratory for Relativistic Astrophysics, Department of Physics, Guangxi University, Nanning 530004, China}

\author{Junzhi Wang}
\affiliation{Laboratory for Relativistic Astrophysics, Department of Physics, Guangxi University, Nanning 530004, China}

\author{Peng Xie}
\affiliation{School of Chemistry and Chemical Engineering, Guangxi University, Nanning 530004, China}

\author{Enwei Liang}
\affiliation{Laboratory for Relativistic Astrophysics, Department of Physics, Guangxi University, Nanning 530004, China}

\author{Zhao Wang}
\email{zw@gxu.edu.cn}
\affiliation{Laboratory for Relativistic Astrophysics, Department of Physics, Guangxi University, Nanning 530004, China}

\begin{abstract}
The density functional theory (DFT) is the most versatile electronic structure method used in quantum chemical calculations, and is increasingly applied in astrochemical research. This mini-review provides an overview of the applications of DFT calculations in understanding the chemistry that occurs in star-forming regions. We survey investigations into the formation of biologically-relevant compounds such as nucleobases in the interstellar medium, and also covers the formation of both achiral and chiral amino acids, as well as biologically-relevant molecules such as sugars, and nitrogen-containing polycyclic aromatic hydrocarbons. Additionally, DFT calculations are used to estimate the potential barriers for chemical reactions in astronomical environments. We conclude by noting several areas that require more research, such as the formation pathways of chiral amino acids, complex sugars and other biologically-important molecules, and the role of environmental factors in the formation of interstellar biomolecules.
\end{abstract}

\keywords{Astrochemistry, Astrobiology, ISM: molecules}

\maketitle

\section{Introduction}           
\subsection{Astronomical detection}
\label{sect:Astronomical}

\begin{figure}[htp]
\centerline{\includegraphics[width=8cm]{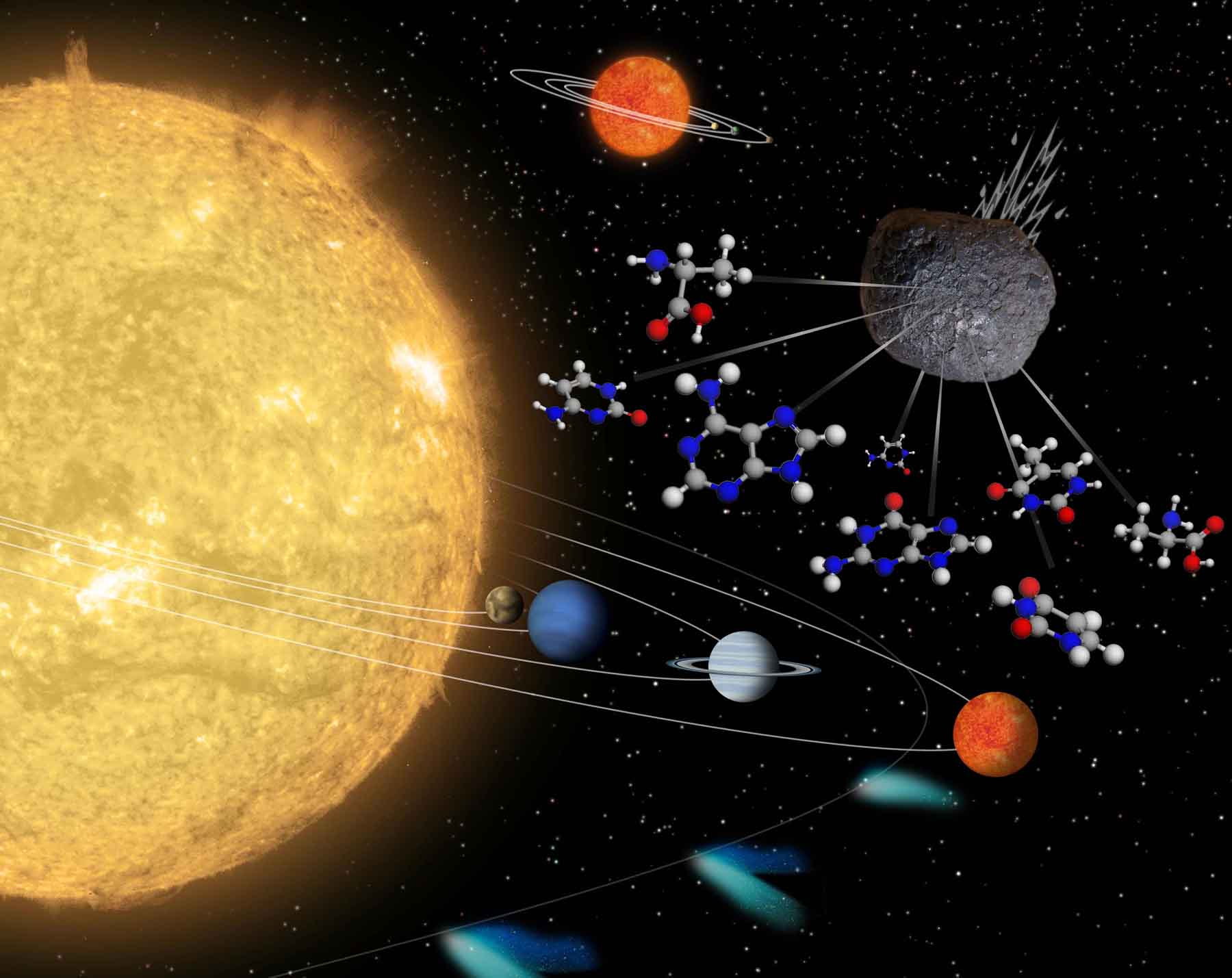}}
\caption{\label{F1}
Schematic for the plausible role of interstellar biomolecules in the emergence of life, through the bombardment of the planets by meteorites in the solar system. The background figure is from the NASA's image library with general permission to reuse.}
\end{figure}

A plethora of biologically-relevant compounds, such as amino acids, nucleobases and sugar derivatives, have been identified in samples of meteorites, comets and asteroids \citep{Burton2012}. In addition, a number of biologically relevant molecules have been observed in the interstellar medium (ISM) \citep{Guelin2022}. These findings strongly suggest an extraterrestrial origin for the elemental building blocks of life on Earth and raise intriguing questions about the formation of these molecules in space and their delivery to planets (as depicted in Figure \ref{F1}). Addressing these questions is crucial for understanding abiogenesis and the search for life in the vast expanse of the Universe. To date, over 200 molecular species have been detected in the interstellar and circumstellar medium \citep{Mcguire2022}. Most of these molecules are organic in nature due to the fact that H, C, N, and O are the most abundant elements in the Universe. Although the majority of these organic molecules are abiotic, the reaction between which presents a vast variety of potential complex organic molecules (COMs) which could be employed by life.

Despite the long-held belief that the space environment is inimical to the formation and survival of COMs, the emergence of astronomical spectroscopy techniques has provided evidence to the contrary \citep{Li2020, Li2021, Lei2022, Lopez-Sepulcre2019}. For instance, COMs were detected in various interstellar regions, some of which demonstrate a high degree of organic chemistry. Examples include formamide (\ce{H2NCHO}) and other amides, which were identified in Sgr B2 and Orion KL \citep{Rubin1971, Suzuki2018}. These molecules contain an amino group, which is analogous to the amino acids found in peptide chains, and are thus considered precursors of the genetic material of life. Another type of astronomically detected compound associated with life is the nitrile, a class of molecules containing one or more nitrile functional groups. These molecules act as raw materials for the formation of ribonucleic acid (RNA). Acrylonitrile (\ce{H2CCHCN}) is an example of a nitrile, which was detected in the formation region of a massive star \citep{Suzuki2018}.

The detection of COMs presents several challenges compared to simpler molecules like \ce{CO}, \ce{CS}, and \ce{HCN}. The criteria for COM detection have been summarized by \citet{Snyder2005}, considering factors such as low abundance, complex energy levels, and line confusions caused by similar rest frequencies. Among COMs, amino acids are of particular interest in the context of the origin of life in the universe, but a reliable detection of amino acids have not been achieved to date. The simplest amino acid, glycine, has been reported as a detection in several sources \citep{Kuan2003}, but it has been pointed out that these lines are not actually from glycine \citep{Snyder2005}. Significant discoveries of new COMs in recent decades include molecules with peptide-like bonds such as formamide \citep{Rubin1971}, acetamide \citep{Hollis2006}, N-methylformamide \citep{Belloche2019}, and propionamide \citep{Li2021}. Chiral molecules like propylene oxide \citep{McGuire2016}, sugar-type molecules like ethylene glycol \citep{Hollis2002}, and branched alkyl molecules like iso-propyl cyanide \citep{Belloche2014} have also been detected, marking important discoveries in the field of astrochemistry. However, fundamental questions about the formation of these COMs in space, their ability to withstand radiation, and their mechanisms of transport to planets or through cosmic history remain largely unexplored. Answers to these questions hold the potential to eventually unveil the origins of life, potentially discover extraterrestrial life, and identify habitable environments for humanity. 

\subsection{Chemistry in ISM}

The reactive routes remain as key questions after the detection of raw materials for the formation of biomolecules in space \citep{Aponte2017, Barone2022, Rimola2022}. Due to the molecular complexity, biomolecules are generally difficult to form through direct collisions between molecules of closed shells in the cold ISM environment. Photochemical and radiation chemical processes are thus often considered to be the primary steps in the synthesis of interstellar COMs \citep{Arumainayagam2019, Hollenbach1999, Castellanos2018}. Laboratory studies of astronomical ice analogues suggest that there may be a complex chemistry on the ice surface exposed to radiation in star-forming regions \citep{Materese2021}. Ice mantles are supposed to concentrate molecules, while the radiation from the young star objects can excite molecules, leading to barrierless reactions. But it could be challenging to generate meaningful quantities of biomolecules on ice mantle, due to the short lifetime of excited states and the photo-instability of the intermediate and product species.

In the process of molecular clouds collapsing to form a star, the temperature of the gas surrounding the young stellar object increases passively due to the protostar's radiation. When the temperature is sufficiently high to evaporate the ice mantle, warm molecule gases are observed at temperatures of $100-300$ K, known as hot cores. These hot cores display a rich organic chemistry with the majority of observed \ce{O-}, \ce{N-}, and \ce{S-} containing COMs detected in regions such as Sgr B2 and Orion KL \citep{Herbst2009, Herbst2014}. Additionally, the protoplanetary disk, where planets are formed, has been a focus of research concerning abiogenesis. However, despite only relatively simple organics having been detected in protoplanetary disks \citep{Oberg2020}, the gas temperature can reach comparable levels to those of hot cores, as the disk temperature increases from 10 to several thousand K from the outer layer to the central forming star \citep{Akimkin2013}. In hot cores and protoplanetary disks, molecules at relatively high density (typically $10^{7}-10^{9}$cm$^{-3}$) in these regions can effectively screen out ultraviolet (UV) radiation, and provides the kinetic energy necessary to overcome the barriers for synthesizing COMs free of photochemical and radiation chemical processes.

\begin{figure}[htp]
\centerline{\includegraphics[width=8cm]{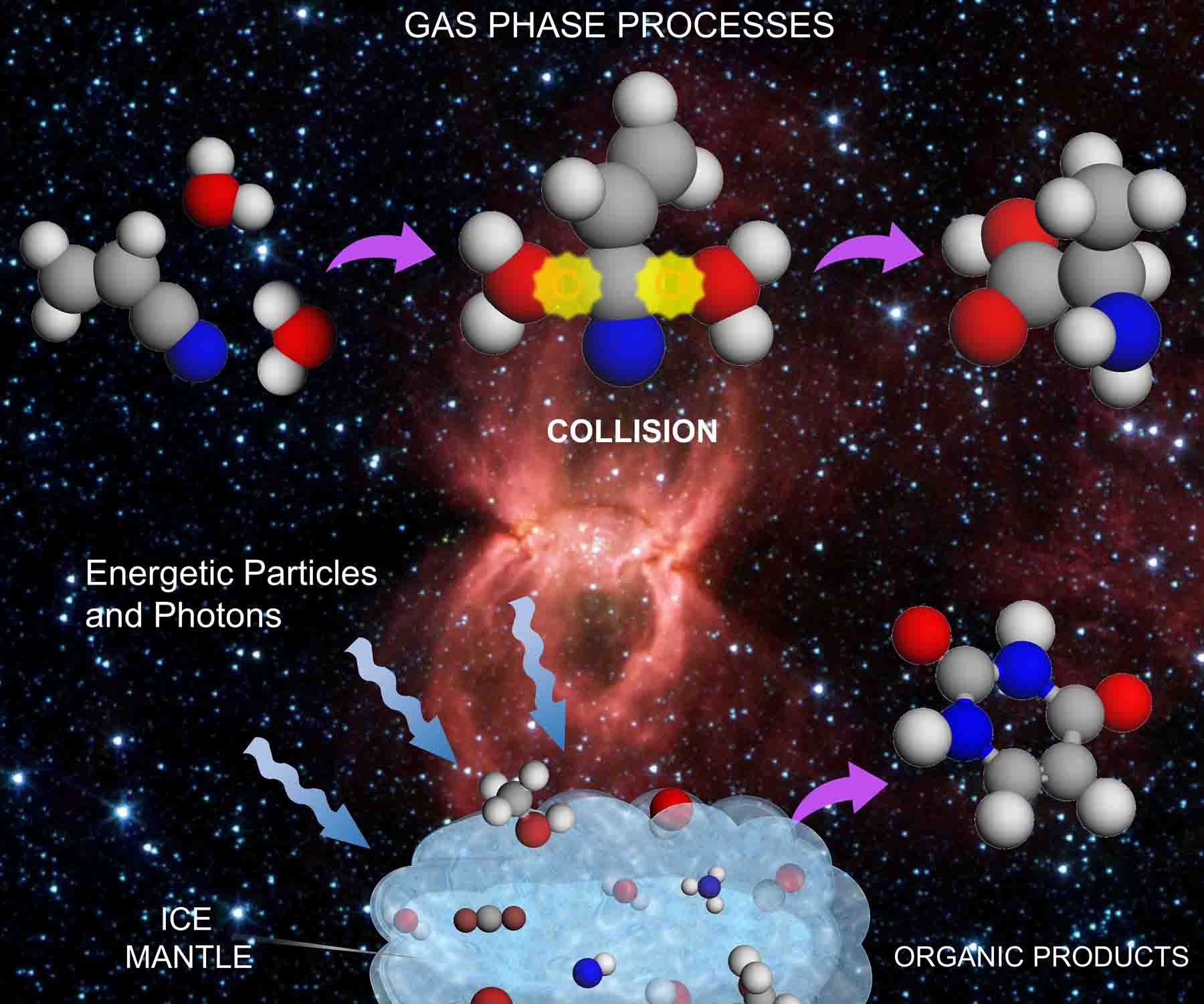}}
\caption{\label{F2}
Schematic for reactions in space occurring in the gas phase (top), and on the surface of grains (bottom). The background figure is from the NASA's image library with general permission to reuse.}
\end{figure}

The chemistry of COMs in the gas-phase interstellar medium is less studied than chemistry on ice by experimental means \citep{Puzzarini2022, Herbst2021, Fuente2019}. There is still debate over the main reaction mechanism that leads to their formation. The majority of COMs with direct biological interests are complex molecules ($\geq 10$ atoms) and the energy barrier for their synthesis from closed-shelled species is usually very high, meaning high temperature needed to activate the process. As a result, previous studies have predominantly focused on ion-molecule reactions or neutral-neutral reactions involving highly reactive radicals \citep{Sandford2020}. However, this viewpoint does not entirely align with the astronomical observation that the majority of interstellar molecules detected so far are neutral species with closed shells \citep{Mcguire2022}.

To this end, researchers have explored catalyzed reactions as an alternative approach to synthesize complex interstellar molecules. However, the limited abundance of suggested catalysts often renders catalyzed reactions seemingly impossible in the ISM, since they require the simultaneous collision of three bodies \citep{Potapov2019,Roy2007,Silva2017,Saladino2013,Qi2018,Hanine2020,Mendoza2004,Rimola2012,Vinogradoff2015}. Fortunately, this is not always the case, particularly when the concentration of the catalyst significantly exceeds that of the reactants, as neutral atomic hydrogen (\ce{HI}). It has been discovered that the presence of \ce{HI} leads to alternative pathways that substantially decrease the energy barriers for reactions, resulting in a significant enhancement of the reaction rate \citep{Yang2023}. Consequently, these reactions become thermodynamically feasible in certain regions of ISM. This suggests that HI-catalyzed reactions could represent a significant mechanism for the interstellar formation of COMs from species with closed shells.

\subsection{Focus of this review}

It is essential to ascertain the intermediates in the formation pathways of interstellar COMs to understand their formation mechanisms. However, many of these intermediates are highly reactive and short-lived, making their detection difficult in laboratory experiments. To this end, chemical networks are developed to model the nature and abundance of interstellar molecules, with the help of spectroscopic observations and laboratory experiments \citep{Unsleber2020, Herbst2013, Zhao2021, Zhang2020a, Zhang2020b, Quan2010}. While this approach has been successful in studying many simple interstellar molecules, it has not been able to explain the observed abundances of COMs. A viable solution to this problem is to perform quantum chemical calculations (QCCs), in which molecular geometry optimization via \textit{ab initio} electronic structure methods can be used to determine the intermediates.

In this concise review, we provide an overview of dedicated QCCs that employ density functional theory (DFT) to model the synthesis of biologically relevant COMs in the ISM, with a primary emphasis on gas-phase chemistry. It's worth noting that previous reviews by \citet{Zamirri2019} have covered the formation of interstellar COMs with a focus on reactions occurring on ice surfaces. \citet{Sandford2020} and \citet{Jorgensen2020} have provided comprehensive reviews of studies concerning interstellar biomolecule formation, primarily focusing on laboratory experiments.

\section{Density functional theory}

QCCs have been utilized in the assessment of electronic contributions to physical and chemical properties of molecules \citep{Sumiya2022, Mata2017}. This involves the systematic application of approximations to the electronic Schr\"{o}dinger equation, with the aim of obtaining electron densities and energies related to the positions of the nuclei and the number of electrons. QCCs are classified according to the type of approximation used to solve the Schr\"{o}dinger equation. The most commonly used approaches are DFT, Hartree-Fock (HF), and their extensions. The HF method is the simplest wave-function based approach. It neglects many-electron correlations and thus has limited accuracy in the description of chemical bond formation. Post-HF methods, such as M{\o}ller-Plesset second order perturbation (MP2) \citep{Moller1934} and coupled cluster single, double, and perturbative-triple electronic excitations (CCSD-(T)) \citep{Gyevi-Nagy2020}, improve the wave function to recover the missing electron correlation, providing increased accuracy, albeit at the cost of reduced computational efficiency. 

The DFT is now widely regarded as the most versatile electronic structure method in QCCs, owing to its high computational efficiency \citep{Jones2015}. In the Kohn-Sham theory, the DFT reduces the many-body problem of interacting electrons in an external field to a problem of noninteracting electrons moving in an effective potential with the exchange and correlation (XC) interactions. Despite its early limitations, DFT has become increasingly accurate for QCCs since the 1990s, following the refinement of its approximations in order to better describe XC interactions.

The Local Density Approximation (LDA) in DFT assumes that the XC energy functional is solely based on the electron density at each point in space. This has been shown to lead to an underestimation of exchange energy and an overestimation of correlation energy. To address this issue, Generalized Gradient Approximations (GGAs) have been proposed, which incorporate higher-order derivatives of the electron density. Despite its widespread use, DFT at the GGA level has been known to have difficulty accurately describing intermolecular interactions, such as van der Waals dispersion, charge transfer excitation and transition states.

The difficulties associated with DFT in describing the exchange interaction can be addressed by the incorporation of a contribution calculated from the HF theory, known as hybrid functionals. An example of this is the popular Becke, 3-parameter, Lee-Yang-Parr (B3LYP) functional, which is based on a linear combination of the HF exact exchange functional and the GGA. An illustrative example of the application of this approach to the abiogenesis question is the determination of the abiotic synthesis pathway of adenine (Ade) from HCN. Experiments have demonstrated that Ade can be synthesized from a solution of HCN and ammonia (\ce{NH3}) since 1960 \citep{Oro1960}. However, the possible pathways of this reaction remained unknown until \citet{Roy2007} performed calculations at the B3LYP DFT level.

It has been suggested that meta-hybrid GGA functionals, such as Minnesota ones, could be potentially more accurate than the popular B3LYP \citep{Peverati2011, Hohenstein2008, Riley2010, Ferrighi2012}. This family includes the M06-L, M06, M06-2X and M06-HF functionals, each with a different amount of exact exchange. Their expansion terms are dependent on the electron density, the gradient of the density and the second derivative of the density. This suite of functionals has demonstrated a successful record for systems containing dispersion forces, thus correcting one of the most prominent insufficiencies of standard DFT methods.

The accuracy of DFT is dependent on the choice of XC functional and the representation of the electronic wave function (namely basis set), which can be composed of either atomic orbitals or plane waves. The former is commonly used for quantum chemistry applications. The most prevalent DFT functional/basis set combinations for astrochemistry include B3LYP/6-31G*, B3LYP/6-311G(d,p), B3LYP/aug-cc-pVTZ, M06-2X/6-31G, M06-2X/6-311G(d,p), and M06-2X/aug-cc-pVTZ, and their derivatives. There are deficiencies associated with the most popular B3LYP/6-31G* model, such as missing London dispersion and basis set superposition error, which have been discussed in \citet{Kruse2012} with regard to DFT calculations of molecular thermochemistry. Additionally, a review by \citet{Mardirossian2017} provides a comparison between different levels of DFT, benchmarking 200 density functionals on a molecular database of almost 5000 data points of non-covalent interactions, isomerisation energies, thermochemistry, and barrier heights. 

Although DFT has proven to be a valuable tool, it possesses certain limitations \citep{Cohen2012}. One significant challenge in studying chemical reactions is accurately describing reaction barriers and energetics. For particular systems, DFT often struggles to predict activation energies with precision, especially for reactions involving complex transition states or significant rearrangements of molecular structures. This limitation arises from the inherent approximation in the exchange-correlation functional, leading to errors in energy barriers. Additionally, handling solvent effects can be challenging, although this is less relevant in the low-density environment of the ISM. To address solvent effects accurately, alternative methods like implicit solvent models, explicit solvent molecules, or hybrid QM/MM models may be necessary \citep{Senn2009, Acevedo2010}. Moreover, DFT encounters difficulties when dealing with reactions involving large systems. Despite being more computationally efficient than most wave-function-based methods, DFT calculations can be demanding for larger molecular systems, such as mixtures containing more than ten molecules. In such cases, molecular dynamics simulations could be more suitable than QCCs \cite{Wang2020, Wang2019, Wang2023}. However, by carefully considering the limitations and implementing appropriate strategies, DFT can still be effectively utilized as a powerful tool for studying interstellar chemical reactions \citep{Jones2015}.

The accurate computation of electron density and free energy of bonding molecules has made DFT a powerful tool for exploring reactive free-energy surfaces. Through the DFT structural optimization approach, it is possible to predict the intermediates and products that constitute a reaction pathway. This process involves updating the positions of the nuclei in colliding molecules towards the minimum of the potential energy, until optimal geometries of the molecules are found at a stationary point on the system's potential energy surface (PES), where the net interatomic force on each atom is acceptably close to zero \citep{Althorpe2003}. To confirm the intermediate and transition state, vibration frequency calculations are usually performed. Additionally, intrinsic reaction coordinate (IRC) calculations can be applied to ensure the transition states reasonably connects the reactants and products. 

The following section will survey DFT studies on the interstellar synthesis of biologically-relevant COMs. The content will be organized by categories of different bio-molecules, including nucleobases, amino acids, sugars and other species.

\section{Nucleobases}

\begin{figure}[htp]
\centerline{\includegraphics[width=9cm]{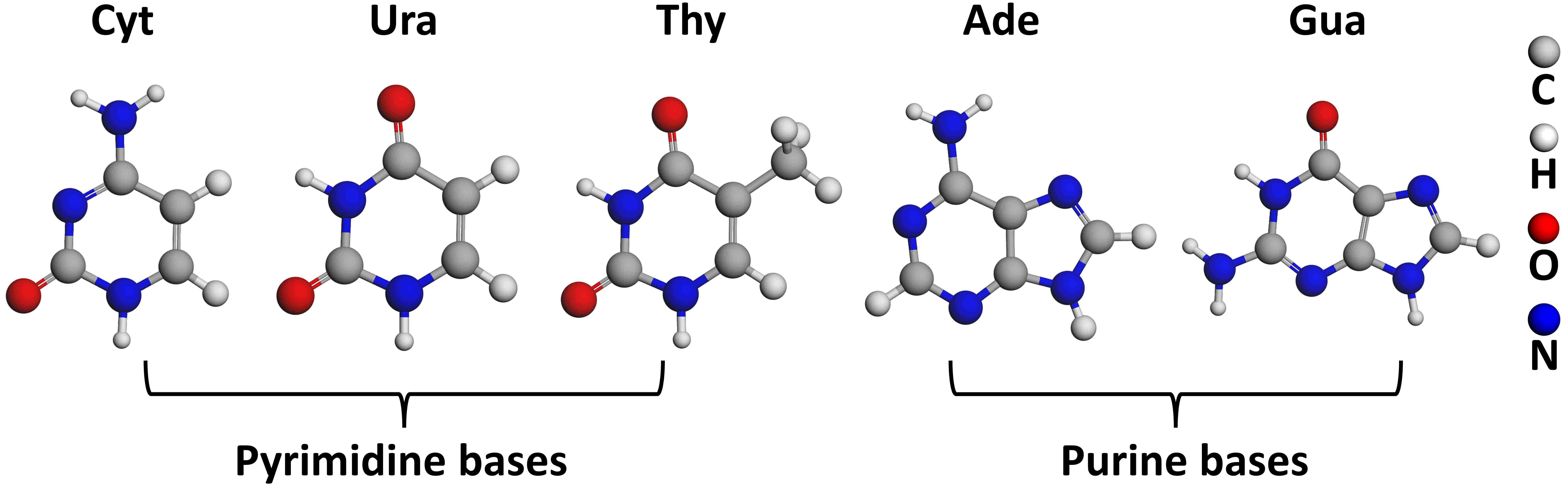}}
\caption{\label{F3}
Chemical structures of the five primary nucleobases.}
\end{figure}

Nucleobases, nitrogen-containing compounds that make up the basic components of nucleic acids, include five primary nucleobases: adenine (Ade, \ce{C5H5N5}), cytosine (Cyt, \ce{C4H5N3O}), guanine (Gua, \ce{C5H5N5O}), thymine (Thy, \ce{C5H6N2O2}), and uracil (Ura, \ce{C4H4N2O2}). The three pyrimidine bases, Cyt, Thy and Ura, are composed of a single ring structure, while the two purine bases, Ade and Gua, have a fused-ring configuration, as shown in Figure \ref{F3}. These five primary nucleobases serve as the fundamental units of the genetic code, with Ade, Gua, Cyt, and Thy present in DNA and Ade, Gua, Cyt, and Ura present in RNA. While direct observations of primary nucleobases in ISM remain elusive, various sets of Ade precursors have been identified. These include aminoacetonitrile \ce{H2NCH2CN} \citep{Belloche2008}, carbodiimide \ce{HNCNH} \citep{McGuire2012}, cyanamide \ce{H2NCH} \citep{Turner1975}, cyanomethanimine \ce{HNCHCN} \citep{Zaleski2013} and glycolonitrile \ce{HOCH2CN} \citep{Zeng2019}. Additionally, hydroxylamine (NH2OH) is a known precursor for pyrimidines and purines \citep{Rivilla2020}, while ethanimine (CH3CHNH) may play a role in amino acid synthesis \citep{Loomis2013}.

\subsection{Pyrimidine bases}

Most studies on the interstellar formation of pyrimidine nucleobases focus on routes from pyrimidine (\ce{C4H4N2}), due to its chemical structure being directly relevant to that of the nucleobases. In the astronomical context, pyrimidine is known to be a precursor of nucleobases in interstellar ice analogs. Using laboratory simulations of interstellar environments, \citet{Nuevo2009} observed the formation of Ura in the residue generated from the UV photolysis of pyrimidine in pure water ices. To explore the mechanism of the reactions, DFT at the B3LYP level was employed to aid in the identification of a number of key intermediates \citep{Bera2010}. \citet{Oba2019} reported the presence of pyrimidine and purine nucleobases in interstellar ice analogs composed of \ce{H2O}, CO, \ce{NH3} and \ce{CH3OH} after exposure to UV irradiation followed by thermal processes. Laboratory experiments demonstrate that the UV photoprocessing of pyrimidine and purine in simple ices of astrophysical interest can result in the formation of all five primary nucleobases \citep{Materese2018}.

\citet{Cole2015} utilized an ion trap mass spectrometer and DFT calculations at the B3LYP level to reveal an interesting connection between the pyrimidine nucleobase anions and cyanate (\ce{OCN-}). Similarly, \citet{Gupta2013} employed the same theoretical method to propose two exothermic formation pathways for Cyt, beginning from propynylidyne (CCCH) and cyanoacetylene (HCCCN). \citet{Choe2021} used DFT calculations to investigate synthesis pathways towards Cyt, Ura, and Thy on icy grain mantles, through reactions between \ce{HCCCN}, protonated isocyanic acid (\ce{H2NCO+}), and one of \ce{NH3}, \ce{H2O}, and \ce{CH3OH}. Furthermore, \citet{Choe2020} proposed a formation route of a Cyt resulting from the reaction of urea with a protonated cyanoacetylene (\ce{CAH+}) and \ce{H2O}.

Pyrimidine has been found to be unstable under UV radiation \citep{Peeters2005}, leading to the consideration of alternative routes for nucleobase formation. Formamide (\ce{H2NCHO}) has been proposed as a reactant in this context \citep{Saladino2016, Rotelli2016}, and it has been demonstrated that the polymerization of its dehydration products can lead to nucleobase formation \citep{Nguyen2015, Jeilani2016}. DFT studies conducted at the B3LYP/6-311++G(d,p) level have suggested that the reactions of free radicals such as CCCNH, CCCO may lead to the formation of pyrimidine bases \citep{Wang2012}. 

Using DFT calculations at the M06/6-31+G(d,p)/6-311++G(d,p) level, \citet{Lu2021} investigate the gas phase reaction between partially dehydrogenated formamide and vinyl cyanide (\ce{H2CCHCN}) for synthesizing Cyt, Thy, and Ura via 1H-pyrimidin-2-one (\ce{C4H4N2O}), which acts as a direct nucleobase precursor. Their results indicate that the H migration is the rate-determining process in the synthesis of the pyrimidine bases. However, it should be noted that highly-dehydrogenated free radicals may be unstable in interstellar molecular clouds due to their high chemical reactivity and consequent short lifetime. Moreover, an interesting approach of automated path search was used by \citet{Komatsu2022} to study the synthesis of Cyt from \ce{NH2CCHO}, based on DFT at the levels of UB3LYP and UM06-2X.

\subsection{Purine bases}

The formation of purine bases Ade and Gua is more complex in comparison to the pyrimidine bases.  Laboratory experiments have employed photochemical reactions of purine in interstellar ice analogues to synthesize Ade, Gua and their derivatives \citep{Materese2017, Materese2018}, with 2-aminopurine and isoguanine hypothesized to be two key intermediates. DFT calculations utilizing the B3LYP functional with the cc-pVDZ basis set have been performed to model this reaction \citep{Bera2017}. The results of the calculations suggest a multistep reaction mechanism involving purine cation, hydroxyl and amino radicals, as well as water and ammonia. The effect of the ice surface was mimicked in the same work using a conductor solvent model.

The pathway of HCN pentamerization has been the focus of many follow-up works. For example, \citet{Gupta2011} used DFT at the B3LYP/6-31G** level to form Ade from HCN, HCCN, \ce{NH2CN}, and CN, and estimated rate coefficient up to $8.71 \times 10^{-9}$cm$^{3}$s$^{-1}$. \citet{Jung2013} used DFT at the B3LYP/6-311G(2d,d,p) level to simulate the Ade synthesis by oligomerizations of HCN, however reporting a high barrier of this reaction (a few hundred kJ/mol). \citet{Cole2015} used a combination of experiments and DFT at B3LYP/6-311++G(d,p) level to study the formation of the purine bases in a reversed manner, by measuring the products of dissociating Ade and Gua. This revealed a number of interesting products, including carbodiimide (HNCNH), cyanamide (\ce{NH2CN}), and isocyanic acid (HNCO). Additionally, \citet{Merz2014} proposed a synthesis pathway towards Ade from \ce{CCCNH} and HNCNH and its isomer \ce{H2NCN} using MP2 calculations.

\citet{Wang2013} proposed a synthesis pathway from formamide to Ade using DFT at the B3LYP/6-311G(d,p) level. Additionally, \citet{Choe2018} utilized DFT at the B3LYP/311G(2d,d,p) level to suggest formation pathways of Gua from 4-aminoimidazole-5-carbonitrile (AICN). In these reactions, \ce{H2O} was considered as a catalyst, leading to substantial decreases in activation barriers. \citet{Jeilani2018} further studied the free radical routes to the formation of Ade and Gua by DFT at the UB3LYP/6-311G(d,p) level, and identified two key intermediates 5-aminoimidazole-4-carboxamide (AICA) and 5-(formylamino) imidazole-4-carboxamide (fAICA). The B3LYP results were compared to M06 data, and it was found that the results were comparable.

\section{Amino acids}

\begin{figure}[htp]
\centerline{\includegraphics[width=7cm]{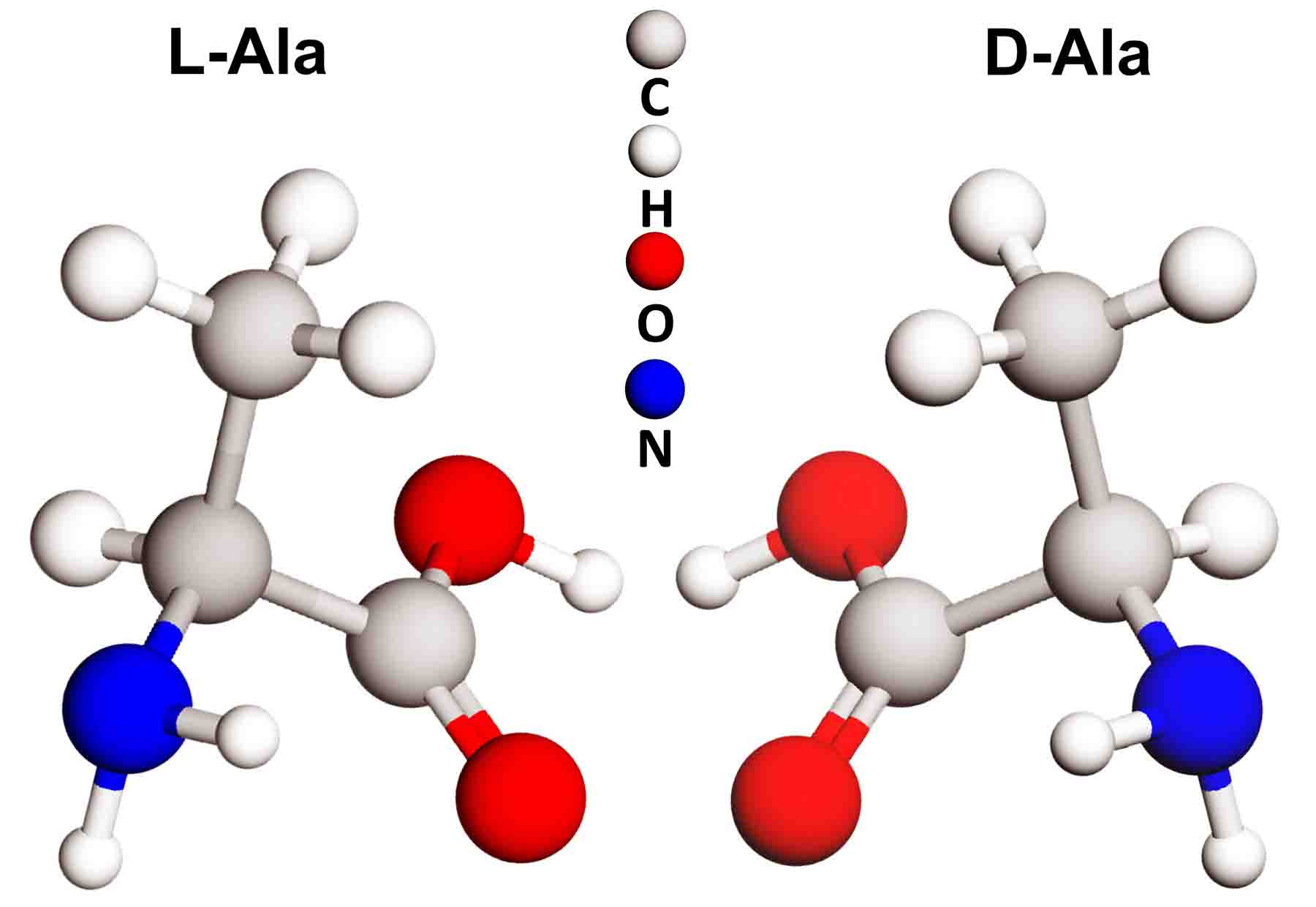}}
\caption{\label{F4}
Molecular representation of the two enantiomers of Ala.}
\end{figure}

Alpha-amino acids, which constitute proteins in the genetic code, have been detected in meteorites and comets, making them of particular interest in astrobiology. All of the alpha amino acids in the genetic code, with the exception of glycine (Gly, \ce{C2H5NO2}), are chiral in nature. That is, these molecules exist in two stereoisomers that are mirror images of each other, referred to as enantiomers. In Figure \ref{F4}, the example of Ala (\ce{C3H7NO2}) is shown, with its two enantiomers: D-Ala and L-Ala, which possess a typical chirality center represented by a central carbon atom bonded to four distinct groups. 

The origin of biological homochirality remains an unresolved question, particularly regarding the chemical processes responsible for the chiral characteristics of alpha-amino acids in the context of astrochemistry. Previous investigations into the interstellar synthesis of amino acids have mainly concentrated on Gly due to its relative simplicity. The proposed mechanisms for its formation often involve Strecker-type synthesis, as well as interactions between radicals and radicals or radicals and neutral species.

\subsection {Glycine}

In \citet{Bernstein2002}, experiments were conducted to explore the formation of Gly and other amino acids when cryogenic water ice containing small amounts of \ce{CH3OH}, \ce{NH3}, and HCN was irradiated with UV light. \citet{Woon2002} evaluated the viability of these pathways by employing QCCs at the MP2 level with aug-cc-pVDZ basis sets, finding that isotopic substitution experiments would identify \ce{CH3OH} as the source of the C in the COOH carboxylic acid group of the amino acids. Subsequently, \citet{Holtom2005} conducted a combined experimental and theoretical study on the formation of Gly and its isomer in interstellar ices, using density DFT at the B3LYP/6-311G(d,p) level. This work revealed that H atoms with sufficient kinetic energy could overcome the entrance barrier to add to a \ce{CO2}, yielding a trans-hydroxycarbonyl radical which then could recombine with other radicals to form Gly and its isomer. Similar studies were undertaken by \citet{Nhlabatsi2016} with DFT approaches, \citet{Rimola2010, Rimola2012} with a cluster approach employing DFT calculations at the B3LYP/6-31+G(d,p) level, \citet{Bossa2010, Oba2015, Sato2018, Thripati2022, Joshi2022} with DFT calculations, \citet{Krasnokutski2020} with DFT+experiments, all proposing pathways to Gly on interstellar ices/grains.

In comparison to surface chemistry, the gas-phase reactions leading to the formation of Gly have been studied to a lesser extent. \citet{Pilling2011} used B3LYP DFT and MP2 calculations to propose formation routes to Gly from carboxylic acids as well as carboxyl radical (COOH), both in the gas-phase and solid-phase. It was suggested that the most favorable gas-phase reactions were between acetic acid (\ce{CH3COOH}) with \ce{NH+} or \ce{NH2OH}, or \ce{NH2CH2} with \ce{COOH+}. \citet{Singh2013} reported that the simple molecules \ce{NH2}, \ce{CH2}, CH and CO could form Gly in both the gaseous and grain phases, based on DFT calculations at the B3LYP/6-31G (d,p) level. \citet{Kayanuma2017} used the B3LYP functional and 6-31G* basis sets to propose synthesis pathways to Gly from aminoacetonitrile (\ce{NH2CH2CN}) and hydantoin (2,4-imidazolidinedione, \ce{C3H4N2O2}) through B\"{u}cherer-Bergs reaction and hydrolysis in both the gas- and solid-phases. In addition, DFT calculations at the B3LYP/6-311++G(2df,p) level is also used to study the formation of Gly peptide chain via unimolecular reactions \citep{Comte2023}. 

\subsection {Chiral alpha-amino acids}

The formation of chiral amino acids in the interstellar medium is not as well understood as that of Gly, despite the fact that homochirality is a common property of amino acids in biosystems. Most of the research in this domain has been focused on relatively simple amino acids such as Ala and Serine (Ser). \citet{Elsila2007} used isotopic labeling techniques to observe multiple reaction pathways for the formation of Gly and Ser in interstellar ice analogs, suggesting that the formation of amino acids is not narrowly dependent on ice composition.

\citet{Shivani2017} suggested, through DFT calculations, that the reaction between \ce{CH3CN}, \ce{CH3CNH2} and HCOOH could lead to the formation of Ala. The same group explored the possibility of Ser formation in the interstellar medium through detected interstellar molecules such as CH, CO, and OH by radical-radical and radical-neutral interactions in the gaseous phase using DFT at the B3LYP/6-311G+(2df,2p) level \citep{Shivani2014}. The results were calibrated with higher-level M06 and M06-2X functional, suggesting that the proposed reactions with low potential barriers could occur in the ISM. \citet{Rani2018} present a study of the pathways of stereoinversion in l-threonine, an amino acid with two stereocenters, employing DFT at the M06-2X/aug-cc-pVTZ level. A simultaneous intramolecular proton and hydrogen atom transfer is observed to drive the stereoinversion, and the reaction rates in the temperature range of $500-1000$ K are found to be significant.

\section{Sugars and other biology-relevant COMs}

The presence of sugars such as ribose and xylose has been identified in meteorites \citep{Smith2020, Laneuville2018}, prompting exploration into the formation of these molecules in star-forming regions. \citet{Woods2013} used DFT calculations to examine the dimerization of the formyl radical and its implications for the formation of glycolaldehyde (\ce{C2H4O2}), the simplest sugar-related molecule. The results suggest that the transition state for dimerization is located within the formyl radical, and that the reaction is energetically favorable under low temperature conditions, consequently leading to the formation of glycolaldehyde. This finding has implications for the formation of monosaccharides, disaccharides, and polysaccharides, all of which are important for astrochemistry and astrobiology.

\citet{Thripati2017} investigated the role of metal ions and hydrogen bonds in the formose reaction, the first step in the synthesis of sugars, utilizing DFT at the B3LYP/aug-cc-pVTZ level. The simulations revealed that metal ions serve as catalysts in the formose reaction, while hydrogen bonds play a crucial role in stabilizing the reaction intermediates. \citet{Vazart2018} and \citet{Skouteris2018} studied the gas-phase formation routes for glycolaldehyde, acetic acid, and formic acid, also known as the ethanol tree, through a combination of infrared spectroscopy and DFT calculations at the B2PLYP/m-aug-cc-pVTZ level. It was observed that radiative association is the principal mechanism for the formation of glycolaldehyde and acetic acid, while gas-phase elimination is the primary route for formic acid. \citet{Jeilani2020} investigated the autocatalytic formose reaction and its potential role in the formation of RNA nucleosides using DFT at the B3LYP/6-311G(d,p) level. Results indicated that the autocatalytic formose reaction is capable of forming several RNA nucleosides, including Cyt and Ura.

A study by \citet{Woon2021} used DFT calculations to investigate the formation of glycolonitrile (\ce{HOCH2CN}) from reactions between C$^{+}$ and HCN and HNC on icy grain mantles. It was demonstrated that the reactions can contribute to the formation of the molecule and that the reaction between C$^{+}$ and HCN is a major contributor in interstellar regions. \citet{Ahmad2020} presented a theoretical approach to elucidate the formation of \ce{C2H4O2} isomers in the ISM through reaction of interstellar formaldehyde molecules, utilizing DFT molecular dynamics simulations. The results revealed that the reaction between formaldehyde molecules is exothermic and occurs in a concerted way to produce three isomers: glycolaldehyde, methyl formate and acetic acid.

DFT calculations have been used to investigate the formation of other biologically-relevant COMs in ISM. \citet{Xie2022} proposed a new pathway in ISM, leading to the formation of biorelevant molecules carrying amine groups or peptide bonds via single-photon ionization induced Michael/cyclization reaction of acrylonitrile (AN)-alcohol heterodimer complexes in the gas phase, combining experiments and DFT calculations. \citet{Ahmadvand2014} investigated the potential energy surface of cyclopropenone (c-\ce{H2C3O}) formation using the B3LYP DFT and found the spin-allowed pathway of c-\ce{H2C3O} formation to be more energetically favorable than the spin-forbidden pathway under interstellar conditions. 

Nitrogen-substituted polycyclic aromatic hydrocarbons (NPAHs) exhibit strong emission features in the infrared spectral region, and are of importance for astronomical observations aiming at the abiogenesis question as they could be building blocks of biomolecules \citep{Meng2021, Kovacs2020, Meng2023}. Parker and coworkers demonstrated the possibility of forming NPAHs from precursor molecules in ISM through a combination of DFT calculations and experiments \citep{Parker2015a, Parker2015b, Parker2015c}. Furthermore, molecular dynamics simulations employing DFT-fitted empirical reactive potentials were used to investigate the formation of interstellar PAHs \citep{Qi2018, Hanine2020}. Finally, DFT calculations at the B3LYP/6-311++G** and 6-31+G** level were used by \citet{Yang2022} to examine the interaction between PAHs and organic molecules for forming PAH-organic molecule clusters. The results suggested that the PAH-organic molecule clusters were formed through hydrogen bonding, with the formation of clusters being more favorable when PAHs were more convex and highly functionalized.

\section{Astronomical implications}

Using DFT, one can calculate the free energy of reactants, intermediates, transition states, or products assuming non-interacting particles at a given thermodynamic temperature \citep{Mcquarrie1999}. For instance, one could use the average temperature of a specific phase of a star-forming region. These calculations can determine the potential barriers and assess the feasibility of a proposed reaction in a given astronomical environment. There are two common strategies for estimating the barrier, each leading to a different value. One approach assumes that the energy does not leave the system by radiation during atomic rearrangements in a reaction, while the other considers that energy dissipates after each step in a pathway. Under the former assumption, the barrier is defined as the difference in free energy between the highest-energy state and the RC, while in the latter, it is defined as the energy needed to overcome the most costly step. Although the true situation lies somewhere between these idealized scenarios, the former approach is typically used for interstellar reactions in the gas phase because the atomic rearrangement is usually faster than the vibration cascade.

\begin{figure}[htpb]
\centerline{\includegraphics[width=9cm]{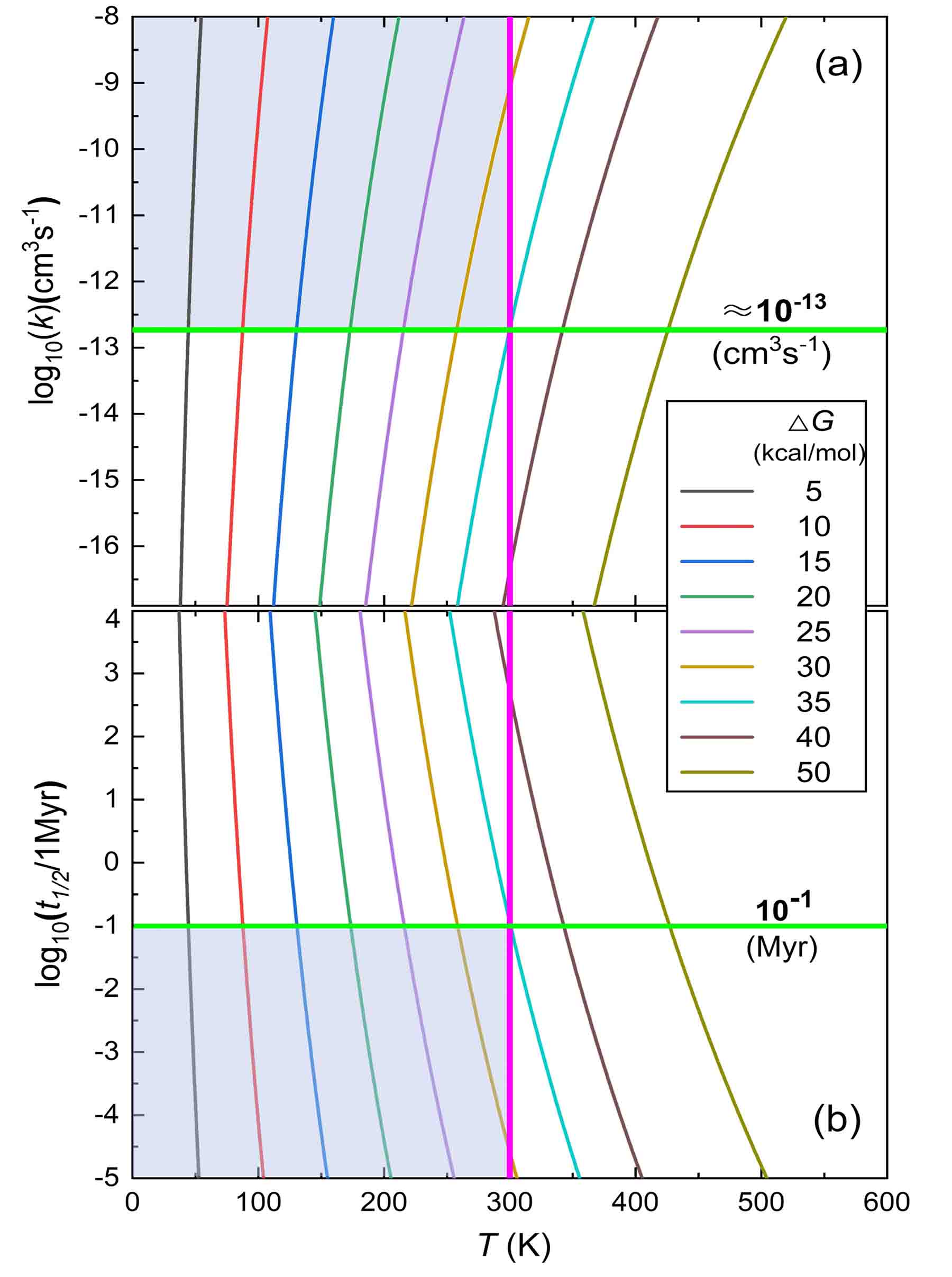}}
\caption{\label{F5}
(a) Reaction rate coefficient $k$ and (b) half-life time $t_{1/2}$ vs temperature $T$ for binary reactions with different potential barriers $\Delta G$.}
\end{figure}

The potential barrier is crucial in determining whether a reaction is feasible in ISM. As temperature increases, molecules collide more frequently and with greater kinetic energy. The proportion of collisions that surpass the barrier increases with temperature, following the Arrhenius relation. The relationship between the reaction rate constant $k$, the half-life time of reactants $t_{1/2}$, the temperature, and the barrier $\Delta G$ is described in the transition state theory \citep{Eyring1935},

\begin{equation}
   t_{1/2}=\frac{ln2}{k}=\frac{h}{C \kappa_{B} T}e^{\frac{\Delta G}{RT}}ln2
   \label{Eq1}
\end{equation}

\noindent where $C$ is the transmission coefficient, $\kappa_{B}$ the Boltzmann constant, $T$ the absolute temperature, $h$ the Planck constant, and $R$ the molar gas constant. The results plotted in Figure \ref{F5} show that, for achieving a half-life time shorter than $10^{5}$ years, a typical astrochemical time frame as highlighted by the green horizontal line in the lower panel, the reaction barrier needs to be lower than ca. $35$ kcal/mol at $300$ K. This limit roughly corresponds to a rate coefficient greater than $10^{-13}$cm$^{3}$s$^{-1}$, as highlighted by the horizontal line in the upper panel. We note that the here-considered reactions are two-body collision processes, or, catalyzed binary reactions in which the catalyst's concentration greatly exceeds that of the two reactants (in the so-called pseudo-binary reactions).

According to Figure \ref{F5}, a potential barrier lower than ca. $35$ kcal/mol signifies a thermally feasibility in the formation regions of young stellar objects, which are often associated with warm ambient gas at average temperatures up to $300$ K, commonly referred to as the hot cores or corinos. These regions typically exhibit a rich organic chemistry and contain the initial materials required for the formation of organic compounds \citep{Herbst2009}. Furthermore, the protoplanetary disk is of particular interest in studying abiogenesis since the chemical composition of the planets formed within it is believed to be inherited from the disk itself. The temperature of the disk gradually increases from $10$ K to several thousand K, ranging from the outer layer to the central forming star \citep{Akimkin2013}.

\begin{table}[h]
\caption{\label{table1} Potential barriers $\Delta G$ (in kcal/mol) of previously-proposed reactions for synthesizing biologically-relevant species in ISM using DFT. Abbreviations for reaction type: r-r (radical-radical), m-m (molecule-molecule), i-m (ion-molecule), r-m (radical-molecule), c (catalyzed).}
\begin{center}

\resizebox{\linewidth}{!}{
\begin{tabular}{ccccc}
\multicolumn{5}{c}{Nucleobases} \\
\hline
    Species & $\Delta G$ & Type & Level of theory & Reference \\
\hline
Cyt & $<12$ & r-r/r-m & \makecell{B3LYP/6-311+G** \\ B3LYP/6-31G**} & \citet{Gupta2013} \\
Gua & barrierless & m-m(c) & B3LYP/6-311G(2d,d,p) & \citet{Choe2018} \\
Cyt/Ura/Thy & barrierless & i-m & B3LYP/6-311G(d,p) & \citet{Choe2020} \\
Cyt & barrierless & i-m & B3LYP/6-311G(d,p) & \citet{Choe2021} \\
Cyt/Ura/Thy & $3-7$ & r-r & B3LYP/6-311G(d,p) & \citet{Nguyen2015} \\
Ade/Pur/Cyt/Ura & $14-37.3$ & r-r/r-m & B3LYP/6-311G(d,p) & \citet{Jeilani2016} \\
Cyt/Ura/Thy & $11.2$ & r-m & B3LYP/6-311++G(d,p) & \citet{Wang2012} \\
Ade & $54-57$ & m-m & B3LYP/6-311G(d,p) & \citet{Benallou2019} \\
Ade & $70.5$ & m-m & B3LYP/6-311G(2d,d,p) & \citet{Jung2013} \\
Ade & $25.7$ & m-m(c) & B3LYP/6-311G(d,p) & \citet{Wang2013} \\
Ade/Gua/Pur & $4.3-29.8$ & r-r & \makecell{UB3LYP/6-311G(d,p) \\ M062x/6-311G(d,p)} & \citet{Jeilani2018}\\
Cyt/Thy/Ura & $>11.3$ & r-r/r-m/m-m & M06/6-31+G(d,p)/6-311++G(d,p) & \citet{Lu2021} \\
Ade/Gua & $>14.1$ & m-m(c) & M06-2X/6-31+G(d,p) & \citet{Yang2023} \\
Ade/Gua/Cyt/Ura & $33.1$ & r-m & B3LYP/6-311G(d,p) & \citet{Jeilani2020} \\

\hline
\multicolumn{5}{c}{Amino acids} \\
\hline
Gly & $71.4$ & m-m & B3LYP/6-31++G(d,p) & \citet{Nhlabatsi2016} \\
Gly & $39.2-40.9$ & r-m & B3LYP/6-31+G(d,p) & \citet{Rimola2010} \\
Gly & $12$ & r-m & BHLYP/6-311++G(d,p) & \citet{Rimola2012} \\
Gly & $1.9-8$ & r-r/r-m & UB3LYP/6-311++G** & \citet{Sato2018} \\
Gly & $18.3-32.9$ & m-m(c) & B3LYP/6-31++G(3df,2pd) & \citet{Krishnan2017} \\
Gly & $15-73.8$ & r-r/r-m & B3LYP/6-31G (d,p) & \citet{Singh2013} \\
Gly & $38-39$ & m-m(c) & B3LYP/6-31G* & \citet{Kayanuma2017} \\
Ala & $2.1-6.8$ & r-r/r-m & B3LYP/6-311G(d,p) & \citet{Shivani2017} \\
Ser & $12-15$ & r-m & B3LYP/6-311G+(2df,2p) & \citet{Shivani2014} \\

\hline
\multicolumn{5}{c}{Sugar and others} \\
\hline
\ce{C2H4O2} & $<89$ & i-m & B3LYP-D3/aug-cc-pVTZ & \citet{Thripati2017} \\
\ce{C2H4O2} & $21.3-23.2$ & r-r/i-m & B3LYP/6-311G(d,p) & \citet{Jeilani2020} \\
Ribose & $25.4-32.7$ & i-m & B3LYP/6-311G(d,p) & \citet{Jeilani2020} \\
\ce{C2H4O2} & $44-95$ & m-m & B3LYP/6–311G(d,p) & \citet{Ahmad2020} \\
c-\ce{H2C3O} & $<45$ & r-m/m-m & B3LYP/cc-pVTZ & \citet{Ahmadvand2014} \\
NPAHs & barrierless & r-m & B3LYP/6-311+G* & \citet{Parker2015c} \\
Pyridine & barrierless & r-m & B3lYP/aug-cc-pVTZ & \citet{Parker2015b} \\
\hline
\end{tabular} }
\end{center}
\end{table}

For synthesizing complex biomolecules from closed-shell species, the potential barrier $\Delta G$ is typically greater than $50$ kcal/mol. However, reactions involving at least one transient species (atoms, radicals, ions) have much lower barriers and are feasible at the temperatures found in star-forming regions. Alternatively, catalyzed reactions have been proposed for the interstellar formation of COMs. Catalysts can facilitate difficult reactions between closed-shell species, making them easier to proceed under laboratory conditions. Various free radicals, ions, acids, and substrates have been suggested as catalysts for the synthesis of interstellar COMs \citep{Potapov2019, Roy2007, Silva2017, Saladino2013, Qi2018, Hanine2020, Jeilani2013, Jeilani2016, Mendoza2004, Rimola2012, Vinogradoff2015}. Notably, neutral atomic hydrogen \ce{HI} has recently been proposed as an efficient catalyst for the interstellar synthesis of Ade and guanine \citep{Yang2023}. As the most abundant free radical in the universe, it is likely that \ce{HI} acts as a general catalyst for the formation of interstellar organics.

\section{Conclusion and perspective}
This mini-review delves into the application of DFT in studying complex interstellar chemistry, particularly for understanding the formation of biologically-relevant compounds. DFT calculations have proven valuable in elucidating the mechanisms behind the formation of these compounds in the ISM, either guiding experimental investigations or predicting reaction pathways. Despite these advancements, numerous unexplored areas remain. In our perspective, future research avenues should encompass the exploration of complex purine base formation, the elucidation of chiral amino acid synthesis pathways, comprehension of intricate sugar and biomolecule production, examination of the influence of metal ions and hydrogen bonding in ISM chemical reactions, refinement of computational models, and assessment of the impact of temperature and environmental variables on interstellar biomolecule formation.



\end{document}